# Weak Matrix Elements on the Lattice — Circa 1995

A. Soni

Physics Department, Brookhaven National Laboratory,Upton, NY  11973

Status of weak matrix elements is reviewed. In particular, $\epsilon'/\epsilon$, $B \to K^*\gamma$, $B_K$, $B_B$ and $B_{B_s}$ are discussed and the overall situation with respect to the lattice effort and some of its phenomenological implications are summarized. For $\epsilon'/\epsilon$ the need for the relevant matrix elements is stressed in view of the forthcoming improved experiments. For some of the operators, (e.g. $O_6$), even bounds on their matrix elements would be very helpful. On $B \to K^*\gamma$, a constant behavior of $T_2$ appears disfavored although dependence of $T_2$ could, of course, be milder than a simple pole. Improved data is badly needed to settle this important issue firmly, especially in view of its ramification for extractions of $V_{td}$ from $B \to \rho\gamma$. On $B_K$, the preliminary result from JLQCD appears to contradict Sharpe *et al.* JLQCD data seems to fit very well to linear $a$ dependence and leads to an appreciably lower value of $B_K$. Four studies of $B_K$ in the "full" ($n_f = 2$) theory indicate very little quenching effects on $B_K$; the full theory value seems to be just a little less than the quenched result. Based on expectations from HQET, analysis of the $B$-parameter ($B_{h\ell}$) for the heavy-light mesons via $B_{h\ell} = \text{constant} + \text{constants}'/m_{h\ell}$ is suggested. A summary of an illustrative sample of hadron matrix elements is given and constraints on CKM parameters (e.g. $V_{td}/V_{ts}$), on the unitarity triangle and on $x_s/x_d$, emerging from the lattice calculations along with experimental results are briefly discussed. In quite a few cases, for the first time, some indication of quenching errors on weak matrix elements are now becoming available.

## Introduction

The Lattice method is now in use for about a dozen years for calculating weak matrix elements [1,2]. Indeed there are by now about a dozen groups (APE, Columbia, FNAL, JLQCD, LANL, MILC, NRQCD, GKPS, UKQCD, Wuppertal...) involved in such efforts. In addition to these many groups there are also several individual attempts. It is abundantly clear that the hadronic matrix elements effort has evolved into a major activity amongst the lattice community. This massive amount of effort, though, is concentrated on a very few types of problems mainly on two and three-point functions. The problems that at first attracted many of us to the lattice, e.g. the $\Delta I = 1/2$ rule, requiring 4-point function calculations, remain essentially unresolved to this day. Fortunately even for the simpler and limited class of problems entailing computations of 2- and 3-point functions we can make important impact to phenomenology. So the efforts are certainly worthwhile.

One extremely attractive ("sexy") and rather unique feature of the weak matrix effort on the lattice, that has been recognized for a long time, is that it can have repercussions far beyond QCD. Since it is difficult to review all the work that is going on anyway and since the primary challenge facing the Particle Physics community at large, at present, is the quest for new physics, I will choose a few calculations below that tend, in my opinion, to enter into considerations that illustrate this feature. Here is the outline:

1) $\epsilon'/\epsilon$;   2) $B \to K^*\gamma$ and related matters;   3) $B_K$;   4) $B_B$;   5) Sample of Hadron Matrix Elements Results;   6)Lattice + Experiment Constraints on the SM;   7) Summary.

For $\epsilon'/\epsilon$ the theoretical underpinning as they pertain to the lattice have been greatly clarified in the last few years through the relentless efforts of Guido Martinelli and collaborators [3]. Unfortunately, there is acute lack of progress in the lattice calculation of the relevant matrix elements. One of the reasons for including this topic in my review is to stress this point and to emphasize that the experimental status implies an important window of opportunity for the lattice.

Four groups [APE [4], UKQCD [5,6], LANL

[7], BHS [8]] have been working on the matrix elements for $B \to K^* \gamma$. In particular UKQCD has recently reported analysis of $B \to K^* \gamma$ as well as of the decays $B \to \pi \ell \nu_\ell$ [9] and $B \to \rho \ell \nu_\ell$ [10] that are all related through Heavy Quark Symmetry (HQS) [11,12].

On $B_K$ new results are being reported using both staggered and Wilson fermions. Perhaps most interesting are those by the JLQCD group [13]. The Columbia [14] group is reporting a relatively high statistics study using dynamical fermions (i.e. $n_f = 2$). Both of these works use staggered fermions. Preliminary results with Wilson fermions from LANL [15], APE [16] and us [17] are also being reported. We have some preliminary results on $B_K$ with $n_f = 2$ "full QCD" configurations using Wilson fermions.

Comparison of all the four unquenched results with their quenched counterparts shows that in each case $B_K^Q$ ($Q \equiv$ Quenched) is just a bit bigger than $B_K^{n_f=2}$.

We [17] are also presenting preliminary results for the $B$-parameters for $B$-mesons i.e. for $\bar{b}s$ and $\bar{b}d$ bound states including an exploratory study using $n_f = 2, \beta = 5.7$ unquenched configurations. Again we found that these parameters are a bit smaller (i.e. just a few per cent at most) in the full theory over their value in the quenched case. Our [17] data also shows clear evidence that $B_{bs}(\equiv B_{B_s})$ is a little (at most a few per cent) bigger than $B_{bd}(\equiv B_B)$.

Finally, I present a sample of the results for hadron matrix elements from the lattice that are of special importance to phenomenology. I then discuss the implications of these lattice results for weak matrix elements for constraining the parameter space of the SM. For this purpose I briefly review the situation with regard to the Wolfenstein parameters $\rho$, $\eta$ as well as $\alpha$, $\beta$, i.e. two of the angles of the unitarity $\Delta$. Numbers emerging for $V_{td}/V_{ts}$ as well as the implications for $B_s - \bar{B}_s$ oscillations are then briefly discussed.

## 1. $\epsilon'/\epsilon$

It is perhaps appropriate to begin the discussion with a brief reminder of the experimental status. After some quarter of a century of inten-

sive efforts the experiments have produced two somewhat conflicting results:

$$|\epsilon'/\epsilon| = (23 \pm 7) \times 10^{-4} \quad \text{NA31 (CERN) [18]}$$
$$|\epsilon'/\epsilon| = (7 \pm 6) \times 10^{-7} \quad \text{E731 (FNAL) [19]} \quad (1)$$

In addition to the disagreement, which is numerically quite significant, the interpretation of the results of the two experiments leads to strikingly different conclusions. E731 being consistent with zero means the Superweak Theory (SW) [20] cannot still be ruled out. On the other hand, if NA31 is correct, not only the SW is clearly ruled out, it is also not inconceivable, as I will stress below, that the result is too high even for the SM and it presents an extremely important hint for physics beyond the SM! The choice between the two alternatives (SM or beyond) is the task of the theorist. It is in this context that lattice calculations have an unusually significant role to play, as will be emphasized in the pages to follow.

In passing it is also worth noting that improved experiments by the two groups are now well on their way on both sides of the Atlantic (NA48 at CERN and E832 ($k$TEV) at FNAL). These are promising to reduce the error significantly, i.e. from about .0007 at present down to .0001!, in the next few years. In particular, it is important to bear in mind that if NA31 is correct then both the experiments should be able to report a clear non-vanishing signal for $\epsilon'/\epsilon$ even before full completion of the experiments i.e. even when the accuracy is only, say, .0002 or .0003 and not the aimed one of .0001. Thus the lattice weak matrix element effort may have a dramatic role to play on the fate of the SM in a few years time.

Recall that $\epsilon'$ is a measure of "direct" CP violation, i.e. occurring at the $\Delta S = 1$ decay vertices as a difference between the amplitudes $\langle \pi\pi | H_W | K^0 \rangle$ and $\langle \pi\pi | H_W | \bar{K}^0 \rangle$. Thus

$$\text{Re} \frac{\epsilon'}{\epsilon} \simeq \frac{1}{6} \left[ 1 - \frac{|\eta_{00}|^2}{|\eta_{+-}|^2} \right] \quad (2)$$

$$\eta_{00} = \frac{\langle \pi^0 \pi^0 | H_W | K_L \rangle}{\langle \pi^0 \pi^0 | H_W | K_S \rangle}$$

$$\eta_{+-} = \frac{\langle \pi^+ \pi^- | H_W | K_L \rangle}{\langle \pi^+ \pi^- | H_W | K_S \rangle}$$

The calculation of the $\Delta S = 1$ effective Hamiltonian to the next-to-leading order (NLO) has been done by two groups [3,21]. The resulting expansion involves the usual ten four quark $LL$, and $LR$ operators.

In evaluation of the matrix elements of the left-right (LR) operators (i.e. $0_{5-8}$) one needs the strange quark mass through, e.g. [3]

$$\langle 0_6 \rangle_{\rm VSA} \sim (f_K - f_\pi) m_K^4 / [m_s(\mu) + m_d(\mu)]^2 \quad (3)$$

Consistent calculations of the matrix elements relevant to $\epsilon'/\epsilon$ thus requires $m_s$ at a scale $\mu$. It should be clear that the lattice approach has the significant advantage that such scale dependent calculations are feasible.

Indeed, considerable progress has already been made in calculating $m_s(\mu)$. In particular, Allton et al. [22] use input from many quenched simulations and to NLO determine:

$$m_s^{\overline{MS}}(\mu = 2 \text{ GeV}) = 128 \pm 18 MeV \quad (4)$$

In this calculation data from various quenched simulations with $6.0 \leq \beta \leq 6.4$ with Wilson and Clover actions is used. From these data $m_s(a)$ as well as its tadpole improved counterpart $\tilde{m}_s(a)$ are obtained through the relations:

$$\begin{aligned} am_s(a) &= \frac{1}{2}\left(\frac{1}{k_s} - \frac{1}{k_c}\right) \\ a\tilde{m}_s(a) &= \ell n\left(4\frac{k_c}{k_s} - 3\right) \end{aligned} \quad (5)$$

These are then related to $m_s^{\overline{MS}}(\mu)$ to NLO accuracy yielding (4).

The error quoted in (4) is claimed to include systematics due to Clover versus Wilson, and due to boosted coupling [23] versus not boosted, $m_s(a)$ versus $\tilde{m}_s a$, and also scale uncertainties. While the error analysis in such a calculation is clearly rather difficult, the range of $128 \pm 30$ appears quite safe.

It is important to note that the scale, $\mu$, most natural for typical lattice computations is $\sim 0(a^{-1})$ i.e. about 2–4 GeV. This is very fortunate as the Wilson coefficients at such a relatively high scale can be systematically calculated by using improved perturbation theory. Indeed at $\mu \lesssim 1$ GeV, which is more suitable for some other non-perturbative methods (e.g. Chiral perturbation theory ($\chi$PT)) Wilson coefficients of some of the important operators (e.g. $0_5$ and $0_6$) are extremely sensitive to the precise numerical value of $\mu$ rendering such methods rather ill suited for reliable calculations [24].

Recall that, in the continuum, by using $\chi$PT, $K \to 2\pi$ matrix elements can be related to those for $K \to \pi$ and $K \to$ vacuum [25]. For lattice computations, this means that the much harder 4-point functions can be obtained as linear combinations of 3- and 2-point functions. Unfortunately, the method cannot be used with Wilson fermions as they do not respect chiral symmetry on the lattice [26]. It does work for the staggered case and in that approach it has become the method of choice [27].

They are several advantages of introducing "$B$-parameters" for these matrix elements, which measure the deviation from vacuum saturation. Thus:

$$B_i = \langle \pi | Q_i | K \rangle / \langle \pi | Q_i | K \rangle_{\rm VSA} \quad (6)$$

Several years ago, the staggered group made some progress in calculating $B_5$ and $B_6$ [28]. They used $\beta = 6.0$ on $24^3 \times 40$ and $16^3 \times 40$ lattices. The results were:

$$B_5, B_6 \sim 1 \pm .1 \quad (7)$$

These calculations, while pioneering, were rather crude. Comparison of the $16^3$ and $24^3$ lattices showed substantial finite size effects [27]; it is therefore not clear how large are volume errors on their bigger lattice although the general expectation is that the $24^3$ lattice should be sufficient. Perhaps a more serious problem is that the beta dependence of $B_5$, $B_6$ as well as of just about every matrix element relevant to $\epsilon'/\epsilon$ has never been studied. The only exception in this regard is $B_K$ which enters $\epsilon'/\epsilon$ indirectly (see below); for $B_K$ the dependence on lattice spacing is quite significant especially for $\beta \leq 6.0$ [29]. So the numbers above (7) must be regarded as rather tentative. Ironically, none of these matrix elements have received any attention in the last 5–6 years.

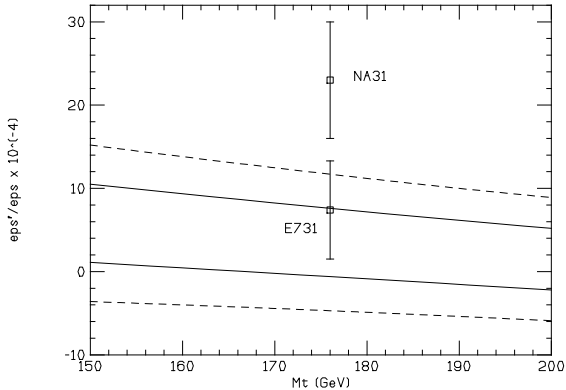

Figure 1. Theoretical expectations for $\epsilon'/\epsilon$ [3,32], (solid 68% CL, dashed 95% CL). Experimental results [18,19] are also shown.

With the improved computational resources that are now available many of these matrix elements should be calculable with appreciable precision at least in the staggered formulation.

With Wilson fermions, in principle, the most attractive method is to attempt $K \to 2\pi$ "directly" [30] (i.e. without reducing it to $K \to \pi$ and $K \to$ vacuum), wherein the use of the discrete symmetry CPS makes the theoretical underpinning very clean. I am quite hopeful that with today's computational resources this 4-point functions method may also work. Indeed, in this regard, the KEK method [31] for calculating multipoint functions seems also to be a promising way to attempt. Finally, the use of the boosted coupling and tadpole improved perturbation theory [23] also tends to improve the chiral behavior of Wilson fermions. Thus the use of $\chi$PT to calculate the reduced 3- and 2-point functions also has a fair chance of working now.

In their study Ciuchini et al. [3] use, based on existing, somewhat crude, lattice calculations and some "guesswork" (the ones with * on them):

$\hat{B}_K = .75 \pm .15$    $\hat{B}_{1,2}^c = 0 - 0.15$    $\hat{B}_{3,4}^* = 1 - 6$
$\hat{B}_{5,6} = 1.0 \pm 0.2$    $\hat{B}_{7-9}^{1/2*} = 1.0 \pm 0.2$    $\hat{B}_{7,8}^{3/2*} = 1.0 \pm 0.2$

(note that hatted quantities are renormalization group invariant). Their key result is [32]:

$$|\epsilon'/\epsilon| = (1.6 \pm 1.9) \times 10^{-4} \qquad (8)$$

How stable is this result to some of the inputs and $B$-parameters? To answer this the following variations were tried:

a) Imposing an $f_B$-cut of $200 \pm 40$ MeV yields [32]

$$|\epsilon'/\epsilon| = (2.0 \pm 2.3) \times 10^{-4} \qquad (9)$$

b) Imposing a different $f_B$-cut of $150 \pm 25$ MeV, motivated by the preliminary MILC results [33] you get [32]:

$$|\epsilon'/\epsilon| = (1.9 \pm 2.1) \times 10^{-4} \qquad (10)$$

c) Following the LANL group's preliminary results [7,15], use of $\hat{B}_7^{3/2} = \hat{B}_9^{3/2} = 0.7 \pm 0.15$ yields [32]:

$$|\epsilon'/\epsilon| = (1.7 \pm 1.9) \times 10^{-4} \qquad (11)$$

Thus one arrives at the conclusion that it is very difficult to get $|\epsilon'/\epsilon| > 10 \times 10^{-4}$ [3]. This, of course, leads to a contradiction (see Fig. 1) with NA31 which is the more "significant" of the two experiments, if you take the experimental errors at their face value.

Here are some possible "loop-holes":

1. $0_6$ (or other matrix elements) suffer significant final state interactions (FSI). In this regard, even bounds or clear demonstration of FSI would be helpful.

2. $m_s(\mu)$ is appreciably lower than the Allton et al. results [22]. This seems unlikely.

3. NA31 could be "off".

4. SM is "off".

While the last conclusion is clearly the most dramatic, it is entirely plausible. In this regard, it is important to note that this is SM with CP violation. As is well known the likelihood of the failure of the SM to account for CP violation is not negligible. After all theoretical investigations tend to strongly suggest that SM cannot account for baryogenesis [34].

It is clear then that there is a significant window of opportunity for the lattice in conjunction with new experiments now in progress that will start to produce improved results in the next 2–3 years, to decide the fate of the SM.

## 2. $B \to K^* \gamma$ and Related Matters

Recall that radiative $B$ decays, exclusive (e.g. $B \to K^* \gamma$, $\rho \gamma \ldots$) or inclusive (i.e. $b \to s\gamma$), are important tests of the SM and they are very sensitive to new physics [35]. Also the dynamics of the exclusive modes (i.e. $B \to \rho(K^*) + \gamma$) is important to understand as it controls extraction of $V_{td}$ from these decays. For this purpose a precise understanding of $SU(3)$ breaking in these decays is also very important.

We recall the CLEO measurements [36]:

$$BR(B \to K^* \gamma) = (4.0 \pm 1.7 \pm 0.8) \times 10^{-5}$$
$$BR(b \to s\gamma) = (2.32 \pm .57 \pm .35) \times 10^{-4} \quad (12)$$

From the ratio of these two $BR$ we get [$R_{K^*} \equiv BR(B \to K^* \gamma)/BR(b \to s\gamma)$]

$$R_{K^*}^{expt} \sim 20 \pm 10\% \quad (13)$$

While these experimental observations are extremely important, their quantitative improvements are needed for a meaningful confrontation with theory.

Recall that the hadronization ratio, $R_{K^*}$, is the quantity of direct importance to the lattice. In this regard, the experimental signal for $B \to \gamma K^*$ need not all have a short distance origin [37,38]. The first point is that the lattice calculations of $B \to \gamma K^*$ is, by construction, the matrix element of a very well defined (short-distance), two-quark operator, $O_7$ [8,39]:

$$O_7 = -em_b \bar{s}_\alpha \sigma^{\mu\nu}_{V+A} b_\alpha F^{\mu\nu}/48\pi^2 \quad (14)$$

Furthermore, in identifying the matrix element of $O_7 : \langle B|O_7|\gamma K^* \rangle$ with the experimental observation of $B \to \gamma K^*$, one is also making a tacit assumption that the spectator model works. In other words one is assuming that the light (spectator) quark in the $B$-meson is not playing any important role in $B \to \gamma K^*$ decay.

There are compelling phenomenological reasons due to $V_{ub}V_{us} << V_{ts}$ for believing that the second assumption above is very safe. (Indeed this is not the case for the related decay $B \to \rho\gamma$, especially for $B^- \to \rho^- \gamma$ [37,38]).

However, the first assumption should not be taken for granted. The point is that experimentally it is known that [40]

$$BR(B \to \psi K^*) = (1.58 \pm .28) \times 10^{-3}$$
$$BR(B \to \psi' K^*) = (1.4 \pm .9) \times 10^{-3} \quad (15)$$

which are both substantially larger than $B \to \gamma K^*$. Quantum mechanics tells us that some fraction of the off-shell $\psi(\psi')$ will contribute to $B \to \gamma K^*$. Such a contamination is an example of "long-distance" contribution. It is extremely difficult to make reliable estimates of such LD contributions. Since the experimental result for $B \to \gamma K^*$ contains all contributions—short and long-distance—whereas the lattice calculation, by construction, is SD—the difference between the two is a measure of the LD contributions.

There are by now 4 lattice groups [4–8] (UKQCD, APE, LANL, BHS) who have been studying $B \to \gamma K^*$ on the lattice. At LAT'94, in his review, Guido Martinelli [1] emphasized that the crucial issue was $q^2$ dependence of the relevant form factors ($T_2$ or $T_1$). The point is that if $T_2$ tends to be constant with $q^2$ then $R_{K^*}$ tends to be about 25–35% whereas if $T_2$ is pole-like then $R_{K^*}$ is appreciably smaller, i.e. about 5–10%. Therefore, it is clearly very important to resolve this issue.

On the lattice separate calculation of the matrix element of the vector and the axial piece of the operator is a better strategy as then one has at hand one very good check, namely:

$$T_2(0) = T_1(0) \quad (16)$$

Thus

$$\langle K^*(\eta, k)|V_\mu|B(p)\rangle = 2\epsilon_{\mu\nu\lambda\sigma}\eta^\nu(k)p^\lambda k^\sigma T_1(q^2)$$

$$\langle K^*(\eta, k)|A_\mu|B(p)\rangle = [\eta_\mu(k)(m_{h\ell}^2 - m_V^2)$$
$$- \eta \cdot q(p+k)_\mu]T_2(q^2)$$
$$+ \eta \cdot q\left[q_\mu - \frac{q^2}{m_{h\ell}^2 - m_V^2}\right.$$
$$\left.(p+k)_\mu\right]T_3(q^2) \qquad (17)$$

We recall that at the end point (i.e. $q^2 = q_{max}^2 = (m_{h\ell} - m_V)^2$) $T_1$ and $T_3$ do not contribute. Besides at $q^2 = q_{max}^2$ both the initial and final mesons are at rest requiring no momentum injection rendering the lattice calculation of $T_2(q_{max}^2)$ very clean. Of course, for $q^2 \neq q_{max}^2$ momentum injection is required. Also although $q^2 = 0$, the point of direct physical interest, is not accessible, current simulation parameters do allow $q^2/m_{h\ell}^2$ to be quite small i.e. $q^2/m_{h\ell}^2 \lesssim 0.1$.

## 2.1. Implications of HQS[4,5,8,9,11,12]

HQS provides very useful scaling laws. Thus:

$$T_2(q_{max}^2) \sim \frac{1}{m_{h\ell}^{1/2}} \quad ; \quad T_1(q_{max}^2) \sim m_{h\ell}^{1/2} \qquad (18)$$

In addition

$$1 - \frac{q_{max}^2}{m_{h\ell}^2} = 1 - \left(\frac{m_{h\ell} - m_V}{m_{h\ell}}\right)^2 \sim \frac{1}{m_{h\ell}} \qquad (19)$$

for $m_V/m_{h\ell} \to 0$. Thus if we parameterize the $q^2$-dependence of form factors as

$$T_{1,2}(q^2) = T_{1,2}(0)/(1 - q^2/m_{h\ell}^2)^n \qquad (20)$$

We see that HQS scaling laws suggest that if $T_2$ tends to be constant (i.e. $n = 0$) then $T_1$ must behave as a simple pole (i.e. $n = 1$). On the other hand, if $T_2(q^2)$ falls as a simple pole (i.e. $n = 1$) then $T_1$ should fall as dipole i.e. $n = 2$.

It must be stressed though that these expectations based on HQS are valid only for $m_V/m_{h\ell} \to 0$. In particular, if the lattice heavy-light mass $\sim$ 2 GeV (especially when the simulations are at $\beta = 6.0$) then $m_{K^*}/m_{h\ell} \sim .5$ which may not be small enough for HQS to have set in.

With regard to constant versus a pole behavior of $T_2(q^2)$ two comments are in order. First it is difficult to justify a constant behavior of the form factor as one is dealing with a bound state to bound state transition. Of course, in principle, an approximately constant behavior could result from the cancellation of the amplitudes over two or more resonances. The second point to bear in mind is that it is easier to distinguish between a pole versus a constant as the heavy-light mass gets heavier. If you consider the ratio:

$$T(q_{max}^2)/T(0) = [1 - q_{max}^2/m_{h\ell}^2]^{-1} \qquad (21)$$

then for $m_{h\ell} = 1.5$ GeV, $q_{max}^2/m_{h\ell}^2 \sim .1$ and $T(q_{max}^2)/T(0) = 1.11$. This means at $m_{h\ell} \sim 1.5$ GeV to distinguish a constant from a pole requires the data to have an accuracy a lot better that 11%. On the other hand if $m_{h\ell}$ is heavier, say 4.5 GeV then $q_{max}^2/m_{h\ell}^2 \sim .6$ and $T(q_{max}^2)/T(0) \simeq 2.7$ which is appreciably different from 1 so that even with a 30% accuracy one may be able to decide between a constant versus a pole behavior.

Table 1 lists some of the characteristics of the four groups and their findings with regard to $q^2$ dependence of form factors. Curiously LANL and APE using $\beta = 6.0$ tend to favor a constant behavior and UKQCD ($\beta = 6.2$) and BHS ($\beta = 6.3$ and 6.0) favor a pole-like behavior for $T_2$. Perhaps this is related to the comments in the preceding paragraph as clearly the heavy-light masses accessible at $\beta = 6.0$ are less than at $\beta = 6.2$ and 6.3.

Perhaps the best strategy for deciding between the two extreme options, i.e. $T_2$ constant versus $T_2$ pole-like, is to examine the ratio

$$R_{const} = T_2(q_{max}^2)/T_1(0) \qquad (22)$$

(Recall that $T_2(0) = T_1(0)$). Using the $\beta = 6.0$ and 6.3 data of Ref. 8 [extracted from Table III of that ref.] one finds that $R_{const}$ tends to be larger than 1 especially for the $\beta = 6.3$ data point. Indeed fitting the data to a constant one finds $R_{const} = 1.39 \pm .14$ i.e. over $2\sigma$ away from 1. Next we examine the same data slightly differently, as:

$$R_{pole} = \left(1 - \frac{q_{max}^2}{m_{h\ell}^2}\right) T_2(q_{max}^2)/T_1(0) \qquad (23)$$

Now numerically one finds, $R_{pole} = .98 \pm .10$. Thus at least this data set tends to favor the simple pole behavior for $T_2$ and tends to disfavor a constant behavior fairly strongly.

Table 1
For $B \to K^* \gamma$, the groups, their lattice parameters and results.

| Group | $\beta$ | Size | # of Configs | Action | $T_2$ tends to favor | $R_K$.% $T_2$-const | $R_K$.% $T_2$-pole |
|---|---|---|---|---|---|---|---|
| LANL [7,15] | 6.0 | $32^3 \times 64$ | 100 | Wilson | const | $27 \pm 3$ | 4–5 |
| APE [4] | 6.0 | $18^3 \times 64$ | 170 | Clover | const | $31 \pm 12$ | $5 \pm 2$ |
| UKQCD [5] | 6.2 | $24^3 \times 48$ | 60 | Clover | pole | $35^{+4}_{-2}$ | $13^{+14}_{-10}$ |
| BHS [8] | 6.3 | $24^3 \times 61$ | 20 | | | | |
| | 6.0 | $24^3 \times 39$ | 16 | Wilson | pole | N/A | $6.0 \pm 1.2 \pm 3.4$ |
| | 6.0 | $16^3 \times 39$ | 19 | | | | |

Recall that HQS tends to relate $B \to K^* \gamma$ form factor to $B \to \pi$ [5,9,11,12].

$$T_2^{B \to K^*} \to f_0^{B \to \pi}$$
$$T_1^{B \to K^*} \to f_+^{B \to \pi} \qquad (24)$$

Similarly form factors for $B \to K^* \gamma$ are related to the axial current form factors for $B \to \rho \ell \nu$ [5,9,11,12]:

$$2 T_2^{B \to K^*} \to A_1^{B \to \rho} \qquad (25)$$

where SU(3) is assumed.

Recently UKQCD has studied form factors for $B \to \pi \ell \nu$ [5] and for $B \to \rho \ell \nu$ [9]. Their data for $B \to \pi \ell \nu$ quite strongly favors a pole (dipole) dependence for $f_0(f_1)$ over a constant (pole) one. Similarly their data for $B \to \rho \ell \nu$ favors a pole behavior over a constant one for $A_1$. Finally FNAL group [41] also has a preliminary result for the $f^+, f^0$ form factors for $B \to \pi \ell \nu$ ($\beta = 5.9$, $16^3 \times 32$, 100 configs.). Their data has very good statistics and in particular a constant behavior for $f_0$ is strongly disfavored.

## 2.2. More HQS tests on form factors

UKQCD [9] has studied the HQS expectations on ratios of form factors. They find that HQS relation $[A_1/2T_2 = 1]$ holds to about 20% accuracy even at $m_D$. However, $V/A_1 = 1$ fails at $m_D$ and works quite poorly even at $m_B$ (at 30% level).

BHS [8] studied the HQS relation

$$\sqrt{m_{h\ell}} T_2(q^2_{\max}) = C_1 + C_2/m_{h\ell} + \cdots \qquad (26)$$

Their data in the range of $2 \lesssim m_{h\ell} \lesssim 4$ GeV gave a reasonably good fit to this relation.

## 2.3. Summary on $B \to K^* \gamma$ and related matters

1. A constant behavior for $T_2$ (or for $A_1$) appears to be disfavored.

2. Of course a dependence milder than implied by the simple pole ansatz for $T_2$ cannot at this point be ruled out. More data is needed to examine this possibility critically.

3. Various tests of HQS seem to work quite well for form factors for $m_{h\ell} \gtrsim m_D$. A notable exception is $V/A_1$.

4. Finally a suggestion for analysis. Since HQS makes a very useful prediction for 3-point function e.g. for $T_2$:

$$\sqrt{m_{h\ell}} T_2(q^2 \max) = \text{Const}_1 + \text{Const}_2/m_{h\ell} + \cdots (27)$$

as well as a similar relation for the two point function i.e. the decay constant $f_{h\ell}$. This suggests that it may be better to analyze the data as ratios (3-point functions over 2-point functions) so that $T_2(q^2_{\max})/f_{h\ell}$ could have improved convergence with respect to $m_{h\ell}$. The additional advantage would be that the lattice should be able to provide more accurate predictions for such ratios.

## 3. $B_K$

We recall that $B_K$ defined as:

$$B_K = \langle K | [\bar{s} \gamma_\mu (1 - \gamma_5) d] | \bar{K} \rangle$$
$$/ \langle K | [\bar{s} \gamma_\mu (1 - \gamma_5) d]^2 | \bar{K} \rangle |_{\text{VSA}} \qquad (28)$$

controls the CP violation in the neutral kaon complex through the parameter $\epsilon$. As is well known this parameter ($\epsilon$) has the unique feature that it is the one and only manifestation of CP violation seen so far. What we can learn from the measured value of $\epsilon$ about the origin of CP violation is limited by how well we know $B_K$.

## 3.1. $B_K$ with Staggered Fermions

At LAT'93 Sharpe [42] reported significant progress in pinning down $B_K$ with staggered fermions:

$$B_K^{\text{NDR}}(2\text{ GeV}) = .616 \pm .020 \pm .017 \qquad (29)$$

This extraction of $B_K$ was based on Sharpe's proof that the lattice spacing errors occur only at $0(a^2)$ and not at $0(a)$. In view of the significance of $B_K$ it is clearly important to continue to scrutinize it with both staggered and with Wilson fermions.

I will report briefly on the following results:

1. JLQCD [13] (staggered, quenched).

2. Columbia group [14] (staggered, "full" QCD i.e. $n_f = 2$).

3. APE [16] (Clover action).

4. LANL [15,7] (Wilson).

5. Bernard and Soni [17] (Wilson, see also Lat'94).

JLQCD group is reporting perhaps the most interesting result on weak matrix elements at this year's symposium. They have data with reasonably high statistics at four values of $\beta$. Their simulation parameters are listed below:

| $\beta$ | Size | # of Configs. | $a^{-1}$ GeV |
|---------|------|---------------|--------------|
| 5.85 | $16^3 \times 32$ | 60 | 1.34(3) |
| 5.93 | $20^3 \times 40$ | 50 | 1.58(4) |
| 6.0 | $24^3 \times 40$ | 50 | 1.88(5) |
| 6.2 | $32^3 \times 64$ | 40 | 2.62(9) |

Fig. (2) shows their key result. Their data for $B_K$ (2 GeV) fits very nicely linearly with $a$ (the lattice spacing) i.e. to the form $A + Ba$. In contrast their data excludes the form $A + Ba^2$! If true this result would contradict Sharpe's proof [42].

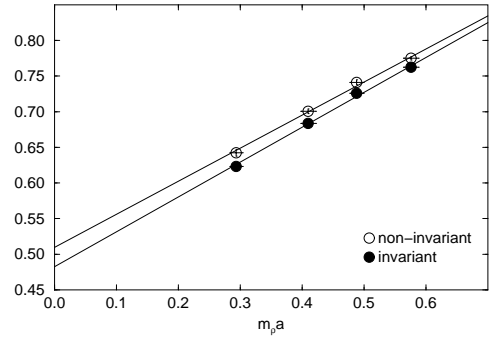

Figure 2. JLQCD fit of $B_K[NDR, 2\,GeV]$ to a linear dependence on the lattice spacing, i.e. $m_\rho a$ for both gauge invariant and non-invariant operators; see also [13].

Fig. (3) shows the JLQCD data along with Sharpe $et$ $al.$ (marked "GKPS") at $\beta = 6.0$, $6.2$ and $6.4$. The GKPS data tends to lie systematically a little bit above the JLQCD one, although, a direct comparison of the two data sets (i.e. without renormalization factor etc.) shows that the discrepancy is quite reduced [13].

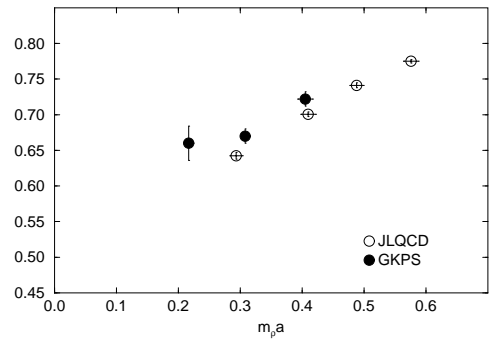

Figure 3. Comparison of JLQCD data for gauge non-invariant operators with Gupta $et$ $al.$ (GKPS) [42]; see also [13]

The most striking point of the JLQCD result [13] is that their data yields such a very high $\chi^2$ ($\sim 25$) to the fit of the form $A + Ba^2$. If their present result withstands further scrutiny it will be a clear contradiction to Sharpe's proof. However, two obvious items that need to be checked before drawing strong conclusions are:

1. Fits of the form $A + Ba^2 + Ca^4$.

2. Perhaps one should discount data at 5.7, 5.9. Then JLQCD would need to run at additional $\beta$'s which they are planning to do anyway. So these questions will be addressed in the next few months. Meantime their *preliminary* result is [13]:

$$\begin{aligned} B_K(2 \text{ GeV}) &= .497 \pm .008(\text{stat}) \\ &\pm .014(0(g^4)) \end{aligned} \tag{30}$$

This central value of $B_K$ is significantly lower than that of Sharpe *et al.* If it is proven correct then it will clearly have very important implications for phenomenology. Also the JLQCD value of $B_K$ tends to be somewhat lower than implied by large $N$ [43]—though not in disagreement (due to the size of the error ($\simeq .05$) that is usually quoted in the large $N$ result) and would, on the other hand, be a little closer to the lowest order chiral perturbation theory value [44].

The Columbia group [14] is reporting a very interesting measurement of $B_K$ in "full" QCD i.e. with $\eta_f = 2$, $am = 0.01$, $\beta = 5.7$ on a $16^3 \times 32$ lattice with 250 configurations. Their key result is:

$$B_K^{\text{NDR}}(\mu = \pi/a) = .659 \pm .063 \tag{31}$$

This result can be compared with the quenched (staggered) one at $\beta = 6.0$

1. Sharpe *et al.* [42] give $B_K^{\text{NDR}}(2 \text{ GeV}) = .707 \pm .008$.

2. JLQCD gives $B_K^{\text{NDR}}(2 \text{ GeV}) \simeq .69 \pm .01$ (this is read off of their figure).

The Columbia study is the third attempt to examine $B_K$ with dynamical quarks using staggered fermions. The previous two were the work of Kilcup [45] and that of Ishizuka *et al.* [46]. In all three $\beta = 5.7$, $n_f = 2$, with the lightest dynamical quark mass $am = 0.01$, is used. All three studies indicate very little difference with the quenched result at $\beta = 6.0$. So the general expectation is that most of the difference between full QCD and QQCD can be accounted for by a shift in the lattice spacing seems to hold to a very good approximation. We will come back to this point a littler later on.

### 3.2. $B_K$ with Wilson Fermions

As is well known, with Wilson fermions, calculation of $B_K$ becomes rather problematic due to the fact that Wilson fermions do not respect chiral symmetry[26,47]. Three approaches are currently being used to address to this problem: 1) The LANL group is calculating the 4-quark matrix elements for several values of momentum transfer [7,15]. The non-chiral part can then be subtracted non-perturbatively. 2) The APE group [16] is using the Clover action and also evaluating the weak coupling corrections using a very interesting non-perturbative method. As Fig. (4) shows this seems to alleviate the problem at least to some extent. 3) We [17,48] use a method wherein the mixing coefficients amongst the 4-quark operators are allowed to vary to systematically restore chiral symmetry.

The LANL [15] group is calculating the matrix elements on their $32^3 \times 64$ $\beta = 6.0$ lattice (150 configurations) at 5 values of momentum-transfer: $p = (0,0,0)$; $(1,0,0)$; $(1,1,0)$; $(1,1,1)$ and $(2,0,0)$. To $0(p^4)$ $\chi$PT predicts [7,15]:

$$\begin{aligned} \langle \bar{K}^0(p_2)|0_{\Delta s=2}|K^0(p_1)\rangle &= \alpha + \beta m_K^2 + \gamma p_1 \cdot p_2 \\ &\quad + \delta_1 m_K^4 + \delta_2 m_K^2 p_1 \cdot p_2 \\ &\quad + \delta_3(p_1 \cdot p_2)^2 \end{aligned} \tag{32}$$

with $p_1 = (m_K, 0, 0, 0)$. Here $\alpha$, $\beta$, $\delta_1$ are pure lattice artifacts due to mixing with the wrong chirality operators. The coefficients $\gamma$, $\delta_2$ and $\delta_3$

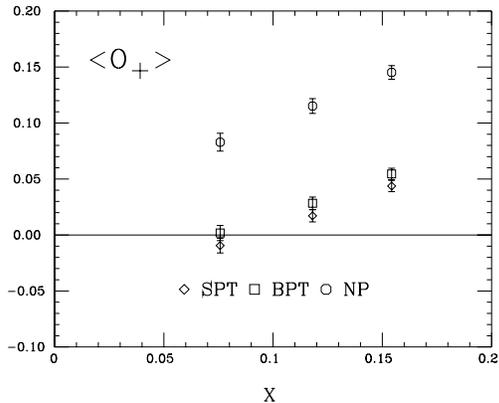

Figure 4. Chiral behavior of the matrix elements of the $\Delta S = 2$ operator $O_+$ as a function of $X = 8/3 f_K^2 M_K^2 / |\langle 0|\bar{s}\gamma_5 d|K\rangle|^2$. Results obtained by renormalisation in "standard" pertubation theory (SPT), in "boosted" perturbation theory (BPT) and non-pertubatively (NPT) are compared; see also [16]

contain lattice artifacts [7,15]:

$$B_K = \frac{3}{8 f_K^2}[\gamma + \delta_2 m_K^2 + \delta_3 m_K(E_i + E_j)] \quad (33)$$

where $E's$ are the energies corresponding to different momenta. After subtractions they find at $\beta = 6.0$ [15]:

$$B_K^{NDR}(2.3 \text{ GeV}) = .74 \pm .10 \quad (34)$$

which is in very good agreement with the staggered result [42]: $.707 \pm .008$.

In our analysis the data set in Table 2 is being used, where $5.7_F$ stands for $\eta_F = 2$, $am = 0.01$ lattices borrowed from the Columbia group [49]. Comparison of $\beta = 5.7_F$ with $\beta = 6.0$ (quenched) $16^3 \times 39$ lattices shows that the errors on $B_K$ due to quenching are small and are within one sigma of the statistical and systematic errors due to other sources. We will take the difference in the two central values (i.e. 0.05) as estimate of quenching errors.

A linear fit (i.e. of the form $A + Ba$) to our data yields:

$$B_K^{a=0}(2 \text{ GeV}) = .58 \pm .06 \pm .01 \pm .05 \quad (35)$$

where the errors are due to 1) statistics and systematics of chiral symmetry restoration, 2) finite size effects and 3) quenching.

### 3.3. Summary of Quenching errors on $B_K$

It is very interesting to note that all four [see Table 3] studies [14,17,45,46] of the effects of quenching (three with staggered and one with Wilson fermions) indicate that effects of fermion loops are small (i.e. $\lesssim 10\%$) and within statistical and systematic errors. However, curiously enough, in all 4 cases the "full" theory result seems to be systematically a little *below* the quenched result. Thus, in each case the ratio $R_{fl} > 1$, where $R_{fl} \equiv \frac{B_K^{n_f=0}(2 \text{ GeV})}{B_K^{n_f=2}(2 \text{ GeV})}$. This may mean that fermion loop reduce $B_K$ by a few %.

### 4. $B_B$: B-parameter for Heavy-Light Mesons

On this topic I will mainly be reporting our own results [17]. I want to take the opportunity to re-iterate that our first study of the $B$-parameter for heavy lights was indeed quite correct [50]. Working at $\beta = 5.7$–$6.1$ although we had used rather heavy quark masses ($am \gtrsim 1$) without using the Kro-Mac norm [52], this does not affect the $B$-parameters as they are ratios in which that norm just cancels. We recall the result of that study:

$$B_{bd}(2 \text{ GeV}) = 1.01 \pm .15 \quad (36)$$

i.e. that for the heavy-light mesons vacuum saturation works to a very good approximation [50,51,53,54]. Indeed since the reduced mass of the heavy-light meson is of the same order as the light meson (e.g. kaon) the agreement with vacuum saturation is non-trivial and was not anticipated. In any case the precise value of the $B$-parameters for the two $B$-mesons ($B_d$ and $B_s$) are extremely important as they enter in constraining the SM CKM parameters through the unitarity $\Delta$ (see the following section).

Analyzing our data we find [17] that $B_{hs}$ (light quark held fixed near the strange quark mass) is increasing with $m_{hs}$ and for $m_{hs} \gtrsim 3$ GeV, $B_{hs}$ is within about 10% of unity i.e. the vacuum saturated value. Note in particular that the data set [see Table 1] contains also the unquenched ("full" QCD) configurations borrowed from the Columbia group [49]. This is the first attempt at studying the difference between the quenched and the unquenched lattices for $B_{h\ell}$.

Table 2
$B_K$ with Wilson fermions using the method Ref. 47 [see also Ref. 17]

| $\beta$ | Size | Configs | $a^{-1}/\text{GeV}$ | $B_K(\mu = 2\text{ GeV})$ |
|---|---|---|---|---|
| 5.7 | $16^3 \times 33$ | 60 | 1.17 | $.81 \pm .03$ |
| 6.0 | $16^3 \times 39$ | 60 | 2.2 | $.66 \pm .08$ |
| 6.0 | $24^3 \times 39$ | 40 | 2.2 | $.67 \pm .07$ |
| 6.3 | $24^3 \times 61$ | 30 | 3.21 | $.59 \pm .07$ |
| 6.5 | $32^3 \times 75$ | 40 | 3.9 | $.70 \pm .05$ |
| $5.7_F$ | $16^3 \times 39$ | 49 | 2.28 | $.61 \pm .07$ |

Table 3
Effects of Quenching on $B_K$

| Study | $R_{fl} \equiv \dfrac{B_K^{n_f=0}(2\text{ GeV})}{B_K^{n_f=2}(2\text{ GeV})}$ | Remarks |
|---|---|---|
| Kilcup [45] | $\sim 1.09 \pm .07$ | Staggered |
| Ishizuka *et al.* [46] | $\sim 1.03 \pm .05$ | Staggered |
| Columbia Group [14] | $\sim 1.06 \pm .10$ | Staggered |
| Bernard & Soni [17] | $\sim 1.08 \pm .15$ | Wilson |

Since gluon interactions with heavy quarks can be expanded in powers of $1/m_{h\ell}$ heavy quark effective field theory suggests that the $B_{h\ell}$ may be expanded in inverse powers of $m_{h\ell}$:

$$B_{h\ell}^\beta = C_0^\beta + C_1^\beta/m_{h\ell} + \cdots \qquad (37)$$

where the superscript $\beta$ is included anticipating the $\beta$-dependence on a lattice calculation. From the lattice perspective, analysis of the data, along such an expansion, should be very useful in attaining precision.

Fig. (5) shows $B_{hs}$ [2 GeV] versus $1/m_{hs}$ from our data [17]. We fit for $C_0^\beta$ and $C_1^\beta$ for each lattice and typically we find

$$C_0^\beta \sim 1.04 \pm .05 \qquad (38)$$

$$C_1^\beta \sim -(.4 \pm .1)\text{ GeV} \qquad (39)$$

In particular, the negative sign of $C_1^\beta$ is unambiguous and furthermore its numerical value of about 400 MeV is characteristic of the mass scale of the "brown muck" [11]. Thus you see that at the $B$-meson mass the $C_1$ term would hardly make a correction of $\sim 10\%$ on the asymptotic value of the $B_{h\ell}$ [2 GeV].

Using such a parameterization we have calculated $B$ for the $B_s$ meson ($M_{bs} = 5.3$ GeV) and we find (see Fig. 3) [17]

$$
\begin{aligned}
B_{bs}(2\text{ GeV}) = {} & .97 \pm .05(\text{statistical}) \pm .01(\text{finite size}) \\
& \pm .02(\text{scale} - \text{breaking}) \pm .02(\text{fitting}) \\
& \pm .04(\text{quenching})
\end{aligned} \qquad (40)
$$

Note again that $B_{hs}$ in "full" QCD (i.e. $n_f = 2$, $\beta = 5.7$, $am = 0.01$) is a bit lower ($\lesssim .05$) then its quenched counterpart at $\beta = 6.0$.

Finally the SU(3) breaking in the $B$-parameters is very important (see below) for relating $B_s$-$\bar{B}_s$ and $B_d$-$\bar{B}_d$ oscillations as well as for extracting $V_{td}$ from the ratio of the oscillation parameters for the two mesons [55]. For such phenomenological applications we have studied the SU(3) breaking ratio:

$$R_{sd} = B_{bs}/B_{bd} \qquad (41)$$

shown in Fig. (6). The following qualitative features are seen:

1. $B_{bs} > B_{bd}$

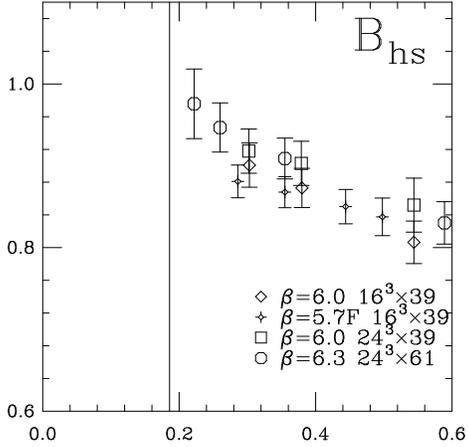

Figure 5. B-parameter for heavy-light (light being roughly at the strange-quark mass) i.e. $B_{hs}$ vs. $1/m_{hs}$; see also [17]. Vertical line indicates the location of the $B_s$ meson.

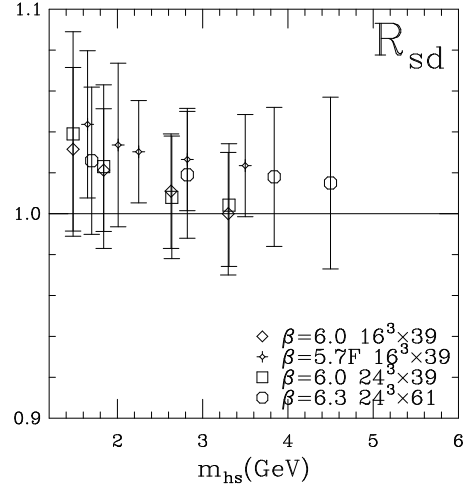

Figure 6. The SU(3) breaking ratio, $R_{sd}$ (i.e. $B_{bs}/B_{bd}$) vs. $m_{hs}$; see also [17]

2. $R_{sd} \to 1$ as $m_{hs} \to \infty$

3. $R_{sd}^{\text{full}} > R_{sd}^{Q}$

Numerically we find

$$R_{sd} = 1.01 \pm .02(st) \pm .02(sy) \pm .03(qu)$$
$$\frac{R_{sd}^{n_f=2}}{R_{sd}^{n_f=0}} = 1.02 \pm .04 \qquad (42)$$

where errors due to statistics, systematics and quenching are as indicated.

## 5. Sample of Hadron Matrix Elements Results from the Lattice.

Table 4 shows an illustrative sample of results for hadron matrix elements obtained from the lattice (recall that $B_{bd} \equiv B_B$, $B_{bs} \equiv B_{B_s}$; these notations are being used interchangeably). Few remarks are in order:

1. On the $B$ paramters since the lattice method is evolving into a precise method for calculating these quantities at a scale $\gtrsim 2$ GeV we are opting to quote their values at

2 GeV. The renormalization group invariant counterpart (often denoted by $\hat{B}_K$) can be obtained from $B_K$ (2 GeV) by multiplying with the Wilson coefficient to a specified order [56].

2. The 90% CL summary that is given for $B_K$ still uses the central value of Sharpe et al. [42] due to the preliminary nature of the JLQCD [13] result; the errors on $B_K$ are increased over those given in previous such summaries [55] to reflect the discrepancy between the two.

3. Errors due to quenching are indicated, in the Table, wherever applicable, by the letter $(Q)$ after the error.

## 6. The Noose: Lattice + Experiment Constraints on the SM [55,65,66].

An important mission of the weak matrix element effort on the lattice, in conjunction with experimental results, is to deduce reliable constraints on the SM. In particular, much attention is focused on the Wolfenstein parameters $\rho$, $\eta$ or equivalently the angles $\alpha$ and $\beta$ of the unitarity triangle [67,40]. From Table 4 using the lattice

Table 4

Illustrative Sample of Hadron Matrix Elements from Lattice QCD

| QUANTITY | VALUE | AUTHORS (REMARKS) |
|---|---|---|
| $B_K$ (2 Gev) | $.616 \pm .020 \pm .017$ | Gupta, Kilcup, Sharpe (Staggered) [42] |
| | $.65 \pm .15$ | ELC (Wilson) [58] |
| | $.58 \pm .06 \pm .05(Q)$ | Bernard, Soni (Wilson) [48,17]. |
| | $.50 \pm .01 \pm .05$ | JLQCD [13,57] |
| $\overline{B_K}$ (2 GeV) | $.62 \pm .15$ | Most likely $\equiv$ 90% CL [59] |
| $B_{bd}$ (2 GeV) | $1.0 \pm .15$ | Bernard *et al.* [50] |
| | $1.16 \pm .07$ | ELC [58] |
| | $.96 \pm .06 \pm .04(Q)$ | B + S [17] |
| $\overline{B_{bd}}$ | $1.0 \pm .15$ | Most Likely (90% CL) [59] |
| $\overline{B_{b_s}/B_{bd}}$ | $1.01 \pm .04$ | B + S [17] |
| $\overline{f_K/f_\pi}$ | $1.08 \pm .03 \pm .08$ | Bernard, Labrenz, Soni [60] |
| $\overline{f_D/}$ (MeV) | $174 \pm 26 \pm 46$ | Bernard, *et al.* [50] |
| | $190 \pm 33$ | Degrand, Loft [61] |
| | $210 \pm 40$ | ELC [58] |
| | $185^{+4+42}_{-3-7}$ | UKQCD [62] |
| | $208 \pm 9 \pm 32$ | BLS [60] |
| | $182 \pm 3 \pm 9 \pm 22(Q)$ | MILC [33] |
| $\overline{f_D/}$ (MeV) | $197 \pm 25$ | Most Likely (90% CL) [59] |
| $\overline{f_{D_S}/}$ (MeV) | $222 \pm 16$ | Degrand, Loft [61] |
| | $234 \pm 46 \pm 55$ | Bernard, *et al.* [50] |
| | $230 \pm 50$ | ELC [58] |
| | $212 \pm 4^{+46}_{-7}$ | UKQCD [62] |
| | $230 \pm 7 \pm 35$ | BLS [60] |
| | $198 \pm 5 \pm 10 \pm 19(Q)$ | MILC [33] |
| $\overline{f_{D_S}/\text{MeV}}$ | $221 \pm 30$ | Most Likely (90% CL) [59] |
| $\overline{f_B/}$ (MeV) | $205 \pm 40$ | ELC [58] |
| | $160 \pm 6^{+53}_{-19}$ | UKQCD [62] |
| | $187 \pm 10 \pm 37$ | BLS [60] |
| | $151 \pm 5 \pm 16 \pm 26(Q)$ | MILC [33] |
| $\overline{f_B/}$ (MeV) | $173 \pm 40$ | Most Likely (90% CL) [59] |
| $\overline{f_{B_S}/}$ (MeV) | $194^{+6+62}_{-5-9}$ | UKQCD [62] |
| | $207 \pm 9 \pm 40$ | BLS [60] |
| | $169 \pm 7 \pm 14 \pm 29(Q)$ | MILC [33] |
| $\overline{f_{B_S}/}$ (MeV) | $201 \pm 40$ | Most Likely (90% CL) [59] |
| $\overline{f_{B_S}/f_B}$ | $1.22^{+.04}_{-.03}$ | UKQCD [62] |
| | $1.11 \pm .02 \pm .05$ | BLS [60] |
| | $1.22 \pm .04 \pm .02$ | FSG [63] |
| | $1.10 \pm .02 \pm .04 \pm .08(Q)$ | MILC [33] |
| $\overline{f_{B_S}/f_B}$ | $1.16 \pm .10$ | Most Likely (90% CL) [59] |
| $R_{K^*} \equiv \frac{\Gamma(B \to \gamma K^*)}{\Gamma(b \to \gamma s)} = 6.0 \pm 1.2 \pm 3.4\%$ | | BHS [8] |
| | $13^{+14}_{-10}\%$ | UKQCD [5,9,64] |
| | $5 \pm 2\%$ | APE [4,64] |
| | $4 - 5\%$ | LANL [7,15,64] |

value of $B_K$ with the kaon CP violation parameter $\epsilon$ leads to one set of curves (see Fig. (7)) which enclose the allowed domain of the SM. Similarly, the $B$ parameter for $B$-mesons and $f_B$ with the experimentally measured $B$-$\bar{B}$ mixing parameter $x_d$, yields two more curves enclosing the allowed area for the SM by this set of considerations. Finally, the experimental observations [40] of semileptonic $B$ decays and the $B$-lifetime yield $\frac{V_{ub}}{V_{cb}}$ (we use $V_{ub}/V_{cb} = .08 \pm .02$ and $V_{cb} = .04 \pm .005$) leading to the two concentric circles on Fig. (7).

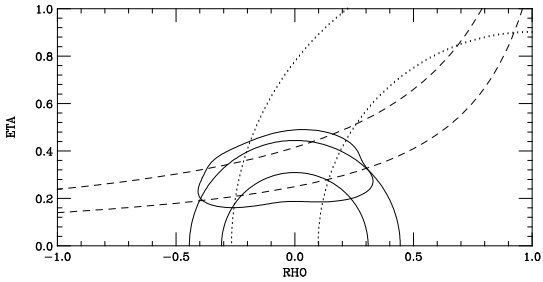

Figure 7. Constraints on the $\rho-\eta$ plane. Concentric circles (solid) are due to $V_{ub}/V_{cb}$; $\epsilon$ and $B_K$ yield dashed curves and $x_d$ with $f_B B_B$ determine the dotted domains. The remaining allowed region (90%CL) is indicated (solid).

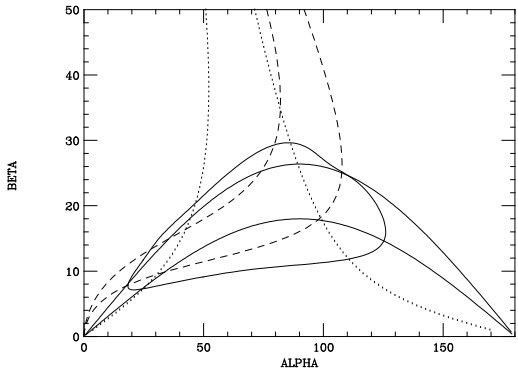

Figure 8. The remaining allowed region (90%CL) in the $\alpha - \beta$ plane (solid). See Fig. (7)

Note that amongst the experimental results providing the SM constraints, so far the lattice has not entered, in any serious way, the deduction of $V_{ub}/V_{cb}$. Of course much work is being done in this context too [68] but the accuracy of the lattice work has not yet reached the same degree of maturity as $B_K$ or $f_B$ [69,33].

Figs. (7-8) show where the SM parameters can still lie. Hopefully improvements in the lattice calculations along with experiments will soon eliminate any allowed domain for the SM in these plots!

Two other important quantities deducible, by use of the lattice results, are $V_{td}/V_{ts}$ and $x_s/x_d$.

Using $f_B = 173 \pm 40$ MeV, $B_B = 1.0 \pm .15$ along with the experimental results [40] on $B_d$-$\bar{B}_d$ mixing ($x_d$) and the measured $B$-lifetime leads to:

$$V_{td}/V_{ts} = .22 \pm .08 \tag{43}$$

Next the relation between the mixing parameters for $B_s$-$\bar{B}_s$ and $B_d$-$\bar{B}_d$ can be quantified.

$$\frac{x_s}{x_d} \sim |f_{B_s}/f_{B_d}|^2 \frac{B_{B_s}}{B_{B_d}} \left| \frac{V_{ts}}{V_{td}} \right|^2 \tag{44}$$

With the ratio $f_{B_s}/f_{B_d} = 1.16 \pm .10$ given before, $R_{sd} \equiv B_{B_s}/B_{B_d} = 1.01 \pm .05$, and the $\frac{V_{td}}{V_{ts}}$ from eqn. (43), we get

$$x_s/x_d \gtrsim 18 \pm 14 \Rightarrow x_s/x_d < 50 \tag{45}$$

Although this ratio is far from precise the upper bound appears safe and already leads to the important experimental implication that $B_s$-$\bar{B}_s$ oscillations may be observable at LEP, SLC and at hadron machines such as HERA-B.

## 7. Summary

1. $\underline{\epsilon'/\epsilon}$: Martinelli *et al.* have clarified the theory as it pertains to the lattice a good deal, however, relevant hadron matrix elements are badly needed. On some cases even bounds could be helpful, e.g. $0_6$. Do these matrix elements suffer from large final state interactions? What is their $a$ dependence? There is a big opportunity here for the lattice to make impact on phenomenology.

2. $\underline{B \to K^*\gamma}$ etc.: Data for "high $m_{ht}$" disfavor a constant $T_2$ (of $f_0$ or $A_1$). Of course, $T_2$ milder than a simple pole cannot, at present be ruled out. It is very important to quantify the $q^2$ dependence more precisely. Improved lattice calculation of $R_{K^*}$ would yield valuable information on "long-distance" contributions. These would have important repercussions for extracting $V_{td}/V_{ts}$ from measurements of $B^0 \to \rho^0 + \gamma$ and $B \to K^* \gamma$ [$B^\pm \to \rho^\pm + \gamma$ is expected to be more "problematic"].

Various HQS relations work very well; not so $V/A_1$.

It may be better to use the lattice data on 3-pt functions and 2-point functions and analyze as ratios of the two. The expected behavior of heavy-light form-factors (at $q^2$ max) and that of the heavy-light decay constant suggests that [form factor (at $q^2$ max) /heavy-light decay constant] may have improved convergence with $m_{h\ell}$ as well as smaller errors.

3. $\underline{B_K}$: JLQCD preliminary result contradicts Sharpe et al. result. In particular it fits very well to $A + Ba$ and not at all to $A + Ba^2$. Needs a lot of scrutiny.

Quenching errors on $B_K$ appear to be very small $\lesssim 5\%$. All 4 studies seem to find, though, $B_K^{full}$ to be just a bit less than $B_K^Q$.

4. $\underline{B_{h\ell}}$ $B_{h\ell}$ should be fit via $B_{h\ell} = $ const $+ \frac{const}{m_{h\ell}}$. Seems to give $B_{bs}(2 \text{ GeV}) = .97 \pm .06 \pm .04$. Again $B_{bs}$ in the full theory seems to be just a little less then in the quenched theory. Also find $B_{bs}/B_{bd}$ is greater that 1 just by a bit.

## Acknowledgements

Special thanks are due to Laura Reina who gave me valuable help in understanding their [3] work on $\epsilon'/\epsilon$ and on updating it. For help in phenomenology and experiments I must also thank David Atwood, Marco Ciuchini, Laurie Littenberg, Giulia Ricciardi, Sheldon Stone, Ed Thorndike and Bruce Winstein. On the lattice side I, of course, received valuable help from Claude Bernard and also from Sinya Aoki, Tanmoy Bhattacharya, Tom Blum, Johnathan Flynn, Brian Gough, Rajan Gupta, Guido Martinelli, Juan Nieves, Masanori Okawa, Steve Sharpe, Jim Simone, Mauro Talevi, Tassos Vladikas and Akira

Ukawa. This research was supported in part under contract number DE-AC02-76CH00016 with the U.S. Department of Energy.